# Comparative Analysis of Hybrid DC Breaker and Assembly HVDC Breaker

Bhaskar Mitra, *Student Member, IEEE* and Badrul Chowdhury, *Senior Member, IEEE*
Department of Electrical and Computer Engineering
University of North Carolina Charlotte
Charlotte, USA
bmitra@uncc.edu b.chowdhury@uncc.edu

*Abstract—* **Voltage Source Converters (VSC) are becoming more common in modern High Voltage DC (HVDC) transmission systems. One of the major challenges in a multi-terminal VSC-HVDC is protection against DC side faults. Two major designs, namely, the hybrid DC breaker and the assembly HVDC breaker, are compared for operational behavior, speed of operation and current carrying capability. The Dual Modular Redundancy (DMR) technique is utilized for decision making of a fault scenario. This uses a dual voting system when one result contradicts the other. This helps in the design of a fail-safe mechanism for the operation of both types of breakers. Current threshold combined with directional change is considered for the breaker operation. A three-terminal bipolar VSC HVDC system is designed in PSCAD/EMTDC and simulation results are utilized to draw a comparison of the two different designs of DC breakers.**

*Index Terms—*Assembly HVDC Breaker, fault, hybrid DC breaker, protection, VSC-HVDC, etc.

## I. Introduction

With the development of off-shore renewable generation from wind, tidal etc., there is a need to make the power available to the on-shore utility grid in by means of overhead lines or cables. A cheap and convenient way of transporting the available generation is through HVDC links, which are able to meet these requirements. Voltage source converter (VSC)-based HVDC systems are gaining importance as it is able to offer a wide range of control and flexibility in power transmission [1]. The VSC-HVDC also provides independent reactive power support at the converter ends, and thus do not require the installation of separate FACTS devices for reactive power compensation [2], [3]. VSC-HVDC systems come with a major drawback - it is highly vulnerable to DC faults. The design of the VSC converter is such that it begins to feed the fault current in the line through the freewheeling diodes in the converter [4], shown in Fig. 1..

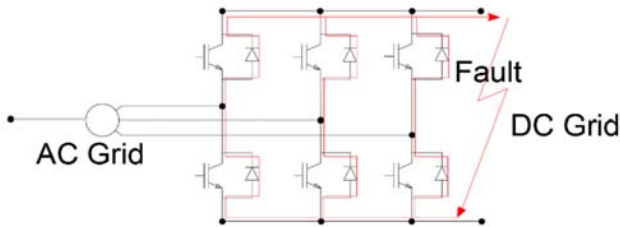

Fig. 1. Freewheeling diode action in VSC-HVDC.

Protection against DC side faults cannot be achieved with the use of traditional AC breakers because DC currents do not have a natural zero crossing as in AC, which can aid in reducing arcing. Besides, conventional AC breakers take on average 200 to 300 ms to interrupt the fault current, which is too slow to protect against faults on the DC system. Specially designed breakers are employed in the DC lines to protect the converter during faults. Due to lack of reactance in the DC system, the rate of rise of the fault current is 5-10 times faster than AC systems and thus the breakers need to operate fast to isolate the system, and prevent a cascaded failure. The fault can be handled by employing a full bridge multi-modular converter (MMC) stations, but this results in a large amount of losses in the system [5], because twice the number of power electronic devices are required, which naturally increases the switching losses and also is costly to implement. The other, more effective method, involves the installation of DC breakers with half bridge MMC submodules. The structure and fault interruption technique of hybrid DC breaker has been discussed in Section II and III. Similarly, in Section IV and V discusses the structure and fault interruption principle of the assembly HVDC breaker. A multi-terminal VSC-HVDC has been modelled in PSCAD/EMTDC, a strategy for their coordination and control has been discussed in Section VI. The simulation results and conclusion on the proposed protection techniques has been discussed respectively in Section VII and Section VIII of this paper.

## II. Structure of Hybrid DC Breaker

Various models of DC breakers have been developed over the years [6], [7], and [8], but they were designed and used for Line-Commutated Converter (LCC) or Current Source Converter (CSC)-based HVDC system, which could break the fault current in typically 30ms~100ms. In addition, LCC systems have a self-fault blocking capability, since the thyristors do not conduct current in the reverse direction. However, for a VSC-HVDC system, this time delay will help the fault current to rise by about 20~25 times, which could resonate through the MTDC grid, and cause a cascading failure. The converter stations lose the control to block the fault current from resonating through the grid, thereby failing to stop it from propagating throughout the grid.

A fault on the DC side of the HVDC grid may be characterized by various stages: a) capacitor discharge phase,

b) diode freewheeling stage and c) grid-side current feeding stage. The diode freewheeling stage is the most dangerous as the diodes carry relatively high current which require fast acting circuit breakers [4].

Semiconductor-based DC breakers have been previously suggested for implementation in HVDC circuits as they are able to interrupt the fault current much faster, but during normal operating conditions, there is a high amount of conduction loss through the breaker, which amount to almost 30%-35% of the total losses in the system [9].

The hybrid DC breaker design, proposed in [10], is similar to that of the solid-state DC breaker but it has two parallel paths namely auxiliary path and main breaker path. The auxiliary path consists of fewer number of semiconductor devices known as load commutation switch (LCS), and thus reduces the loss during normal operating conditions. An ultra-fast disconnect (UFD) mechanical switch, connected in series with the auxiliary unit, helps to transfer the fault current into the main breaker and acts as a protective unit for the LCS during re-connection. The metal oxide varistors are connected in parallel to the main breaker to absorb the excess energy. A detailed schematic of the hybrid DC breaker is shown in Fig. 2.

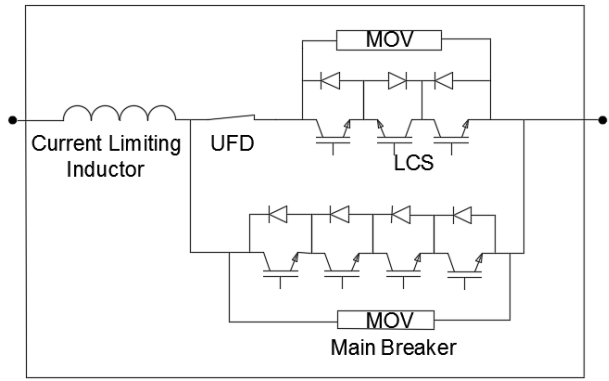

Fig. 2: Hybrid DC Breaker.

### III. FAULT INTERRUPTION IN A HYBRID DC BREAKER

The fault interruption is a hybrid DC breaker occurs in sequential steps, during normal operating conditions the current flows through the LCS and the UFD as it is the path for least resistance as shown in Fig. 3.

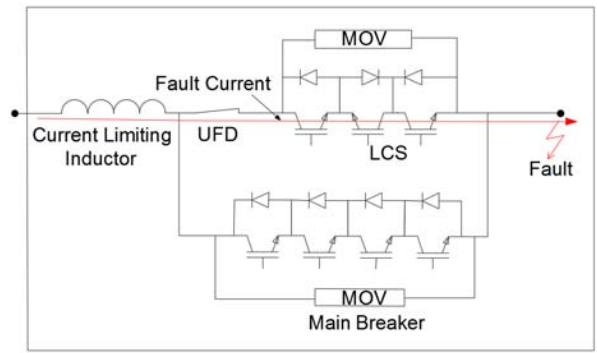

Fig. 3: Normal operating conditions.

At a certain threshold of the line current, the LCS is turned off, and the current is transferred into the main breaker. The UFD opens at zero current in the auxiliary path. This prevents any kind of arcing, and also enhances the longevity of the mechanical switch. The fault current is then transferred into the main breaker, where it is completely interrupted and any excess energy is absorbed in the metal oxide varistors connected in parallel to the main breaker unit. The detailed sequence of operation is shown in Fig. 4 and 5.

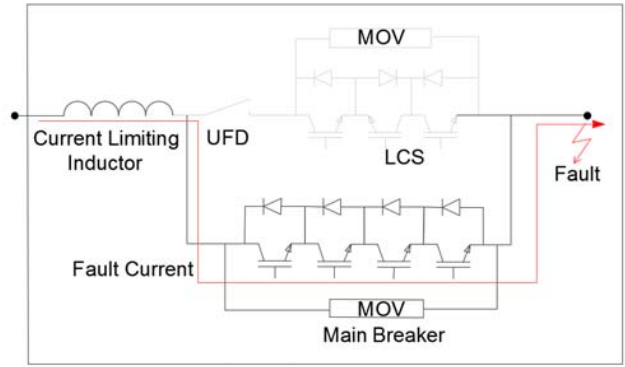

Fig. 4: Fault current transferred to the main breaker branch.

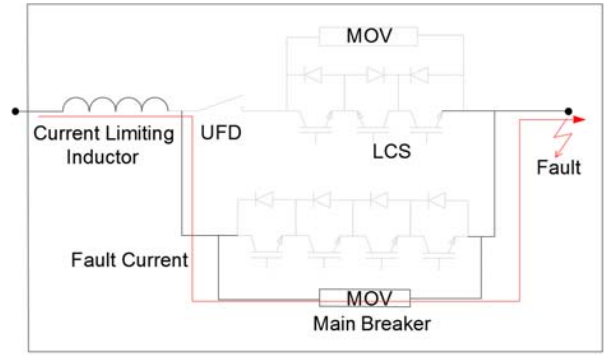

Fig. 5: Excess energy absorbed by the MOV after main breaker operation.

### IV. ASSEMBLY HVDC BREAKER DESIGN

The main breaker section of a hybrid DC breaker, as described in the previous section, comprises of a large number of IGBT devices connected in series since the final fault current interruption takes place here. For a multi-terminal HVDC system, two such hybrid DC breakers must be installed in every transmission line for maintaining reliable operation. Consequently, the number of IGBT devices required are increased, which results in increased installation cost.

The major drawback with regards to the cost of the DC breakers can be overcome by implementing the design proposed in [11], namely, the Assembly HVDC breaker. This also performs the same fault clearing capability, but potentially, at a lower cost. The structure of the assembly HVDC breaker is described below.

#### A. Active Short Circuit Breaker (ASCB)

This is a major component for the assembly HVDC breaker. It is connected in front of every converter station. The breaker is designed to withstand the maximum line-to-line voltage during a fault. Under normal operating conditions, the ASCB remains open. As the fault is detected, gate signal are provided to the IGBT devices to turn-on, whence it creates a shunted branch across the device and reduces the fault current flowing into the system. This allows the main breaker to operate under lower current conditions during a fault.

#### B. Main Breaker

The main breaker has features similar to the load commutation switch (LCS) of the hybrid DC breaker. The number of IGBT modules connected is much fewer as compared to the ASCB unit. The design of the assembly HVDC breaker allows the main breaker design to withstand higher operating voltage which reduces the conduction loss in the system. During normal operating conditions, the device remains operational; it is turned off once the fault is detected and the ASCB has been turned on. This reduces the voltage stress across the device.

*C. Fast Disconnect Switch*

The fast disconnect switch is a mechanical switch which has an operational time of about 2ms~3ms. This switch is a low resistance switch and it remains operational during normal operation. This acts as an isolator switch for the breaker unit. It serves as a protection unit for the main breaker as it reduces the voltage stress across the breaker during initial start. It is normally operated under low fault currents to reduce the arcing and, thereby, increase the lifetime of the switch.

*D. Accessory Discharge Switch (ADS)*

The main function of the accessory discharge switch is that it reduces the voltage stress across the main breaker unit. During normal operation, this remains turned off, but when a fault is detected, the ADS along with the ASCB is ordered to close. The ADS is made up of a series of thyristor switches connected together. The thyristor switches can be employed as the ADS is not designed to interrupt any fault current, rather it acts as a shielding device for the main breaker unit. A resistor unit is connected with the device that helps to discharge the fault current across the resistor helping in the turn-off of the thyristor devices.

A detailed diagram explaining the various components of the assembly HVDC breaker is shown in Fig. 6.

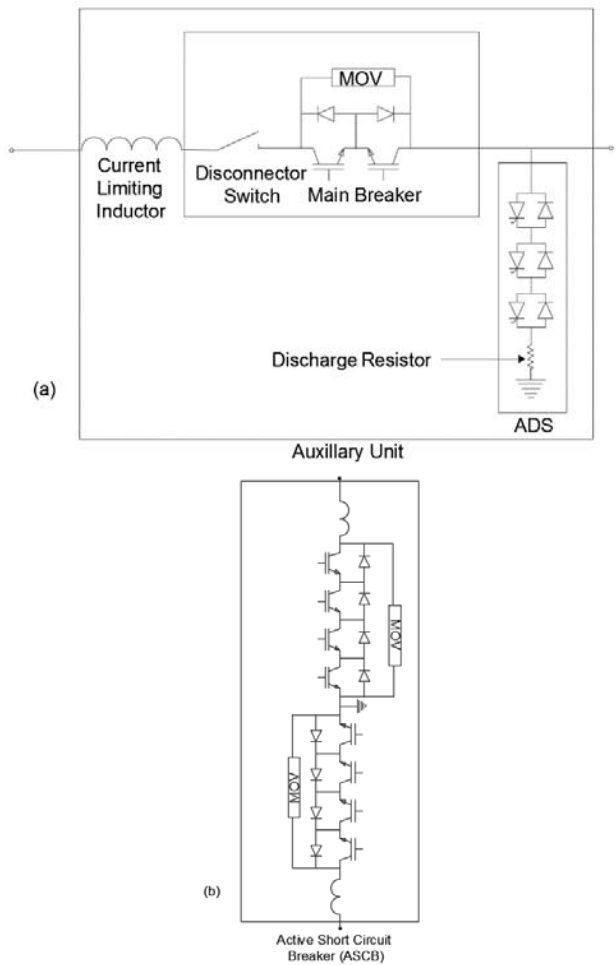

Fig. 6: Assembly HVDC Breaker, (a) Auxiliary unit consisting of the ADS; disconnector switch and main breaker; (b) Active short circuit breaker.

Over the years, various modules of DC breakers have been developed and tested as discussed in [6], [7] and [8]. The highlights are shown in Fig. 7. A summary of their operational characteristics is compared with the Hybrid and Assembly DC breakers in Table I.

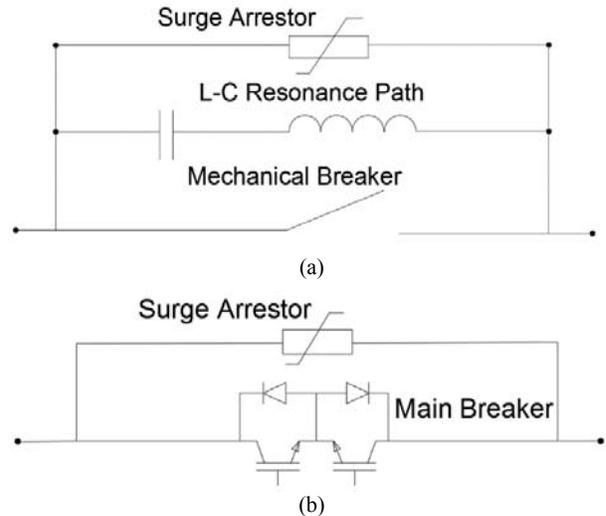

Fig. 7: DC Breakers, (a) Resonant DC Breaker; (b) Solid State Breaker [12].

Table I:
Types of DC Breakers

| | Solid State | Resonant | Hybrid DC Breaker | Assembly HVDC Breaker |
|---|---|---|---|---|
| Commutation [ms] | switch: 0.1 | breaker < 20; resonance: ≤ 30 | switch: 0.1; breaker ≤ 20; UFD 1-4 | main breaker : 0.2; UFD: 2-3; ASCB: 0.3 |
| Interruption time [ms] | < 1 | < 60 | 3-5 | 3-5 |
| Max. Voltage [kV] | 800 | 550 | 750 | 800 |
| Max. Current [kA] | ≤ 5 | 4 | 15 | 15-20 |
| Losses % | 30%-40% | negligible | negligible | negligible |

V. FAULT ISOLATION USING ASSEMBLY HVDC BREAKER

As discussed previously, the assembly HVDC breaker is split into two major components – the ASCB and the auxiliary unit. The ASCB, which is connected at the end of each converter station, is the component where the major interruption of the fault current occurs. The auxiliary unit is connected in series on every line and contains the main breaker. During normal operating condition, the ASCB and ADS remains non-operational. As the fault is detected by the breaker control units, signals are sent to the ASCB and the ADS to close, which creates a temporary short circuit at the DC bus. The flow of the fault current is restricted in the line and this allows the main breaker and disconnect switch to operate at a lower fault current.

The above process enables the main breaker unit to be designed for low voltage operation resulting in lower losses during normal operating conditions. The turning on of the ASCB creates a temporary short circuit at the DC bus which results in the DC bus voltage reducing to almost zero. The operation of the ASCB lasts for about 1ms~2ms which is acceptable for HVDC networks.

After the isolation of the fault is complete, the DC bus voltage is recovered by the opening of the ASCB and the ADS. The closing of the ASCB allows the main breaker to operate at a low fault current condition, and the disconnector switch is opened after a certain time delay. This allows the disconnector switch which acts as a protecting device for the main breaker to operate at zero current conditions, thereby reducing the arcing amount.

The ADS is also designed as a protecting unit for the main breaker. It does not require any switching operations, and thus high frequency switching thyristor devices are not required for its operation. It helps to reduce the voltage stress across the main breaker when it is operated to interrupt the fault current. Thyristor devices, which are durable in nature, are generally installed. The discharge resistor connected in series helps to discharge the current through the ADS to zero, and the thyristors are turned off.

As the ADS are not required to interrupt the fault current but only works as a protective unit for the breaker, they are generally durable. A detailed representation of the step-by step fault isolation in the assembly HVDC breaker is shown in Fig. 8 (a-c).

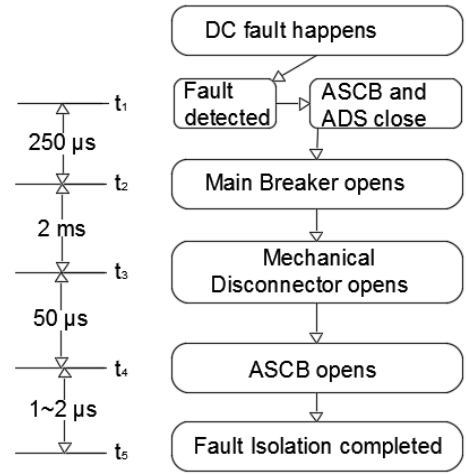

Fig. 9: Fault isolation using Assembly HVDC Breaker.

## VI. FAULT ISOLATION SCHEMES

Various strategies for fault isolation has been discussed in the literature, which involves the use of current threshold [13], ROCOV [14], change in line impedance [15], reactor voltage change rate [16], etc.

A majority of the breakers use the current threshold for the relay units to initiate the breakers, although this is not a fail-safe mechanism. The change in line currents may result in a false indication for the breakers to trigger.

Dual Modular Redundancy (DMR) [17] is a technique which helps to design a fault tolerant mechanism in providing redundancy in case there is a failure of any one of the systems. This technique is used to design a fail-safe mechanism. The current threshold along with the change in direction of the line currents are considered for this scenario.

There is a change in the direction of the flow of current in the faulted line. This change in direction is not observed in the other lines which are not affected. The directional change is shown in Fig. 10.

Any errors with respect to detection and coordination can be avoided by using the DMR technique. The output that is generated from the DMR technique is then compared to that of a voting unit. Only when both the scenarios are in compliance with each other, a decision is reached with regards to the fault detection. An optimized flow chart for fault detection is shown in Fig. 11. The other advantage that is achieved using this technique is that it is not dependent on any form of communication channel, development of which can be a challenge when we want to deploy HVDC systems for remote off-shore generating stations. The breakers are controlled individually, and a failure of a certain unit does not affect the operation of the other breaker.

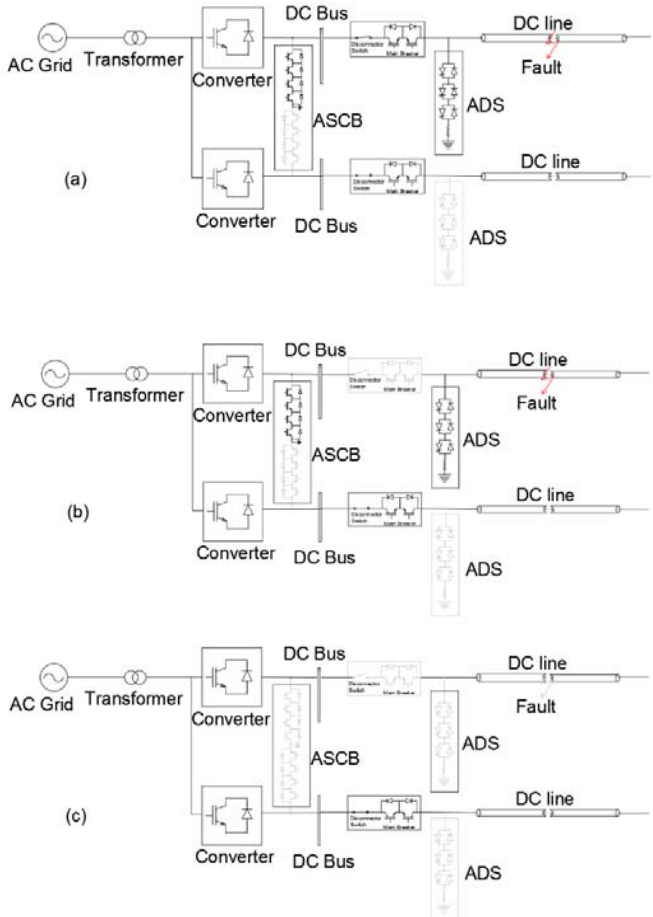

Fig. 8: Sequential operation for fault isolation, (a) the ASCB and ADS are turned on; (b) Main breaker and disconnect switch is turned off; (c) the ASCB and ADS opens completing the fault isolation.

The working principle of the ASCB in fault isolation is provided in the flowchart shown in Fig. 9. The interval t1-t2 is the delay between the fault detection and the opening of the main breaker. The closing of the ASCB creates a temporary short circuit at the DC bus causing the voltage to drop, but this is a temporary operation and since the other lines remains operational, this is not a major concern for HVDC systems.

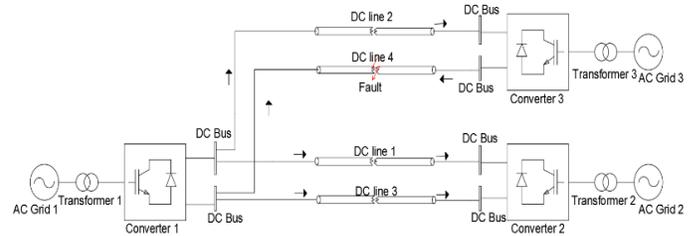

Fig. 10: Change in current direction on a fault cable.

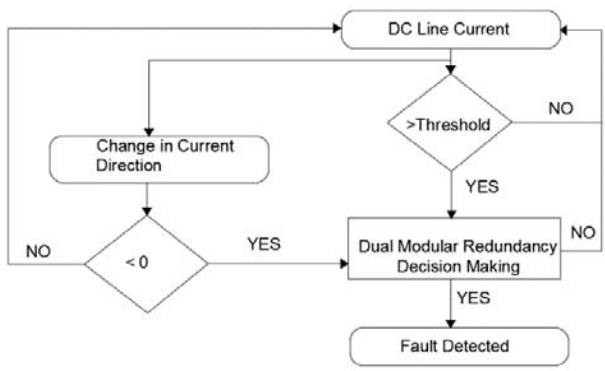

Fig. 11: Fault detection using Dual Modular Redundancy.

## VII. SIMULATION RESULTS

A three terminal default model from PSCAD/EMTDC library is chosen for the studies as shown in Fig. 10. The system has three AC power networks connected to HVDC links using AC-DC converter stations. It is assumed that one of the AC networks is the offshore generating unit connected to two other onshore utility grid networks.

Converter station 1 is considered as the offshore generating unit delivering 150 MW into the DC link, converter 2 is receiving 200 MW which is supplied by converters 1 and 3. Converter 3 also supplies for the station losses. The DC link voltage is maintained at ±420 kV.

A fault was created at t=1.5s at DC line 4, with the system operating at full capacity. The *di/dt* change is much faster than conventional AC systems due to the lack of reactance, and other criteria discussed previously. Therefore, the fault current has the capability to rise 3-5 times of its initial value within 10ms.

There is a threshold limit that is utilized in operating the breakers along with the change in current direction. It can be seen in Fig. 12 that the fault current is interrupted at 1.5kA, and the speed of interruption is within 3-5ms. This is achieved using both the conventional hybrid HVDC breaker and also the Assembly HVDC breaker.

Due to the nature of design and operation of the assembly HVDC breaker, the ASCB is turned on at the converter ends which reduces the amount of fault current flowing through the disconnect switch and the main breaker unit. The fault current interruption takes place in the ASCB which bears the bulk of the stress during its operation. It can be seen in Fig. 13 that the main breaker is operated at a much lower current. Similarly, in the case of a hybrid HVDC breaker, the load commutation switch acts as a transferring branch for the fault current. As a result, the fault current flows through the LCS for a certain time period.

The performance of the hybrid DC circuit breaker and the assembly HVDC breaker are illustrated in Fig. 13 and Fig. 14 respectively. It can be seen that the fault current rises rapidly in both the breakers, but the current in the main breaker unit is arrested as the ASCB is turned on, which prevents the excess flow of fault current. On the other hand, for the hybrid DC breaker, the LCS is turned off only after a certain threshold value of the fault current is reached, which allows the fault current to flow through it for a period of time. It is transferred finally into the main breaker section for final interruption. A comparative analysis of both the breakers is shown in Table II.

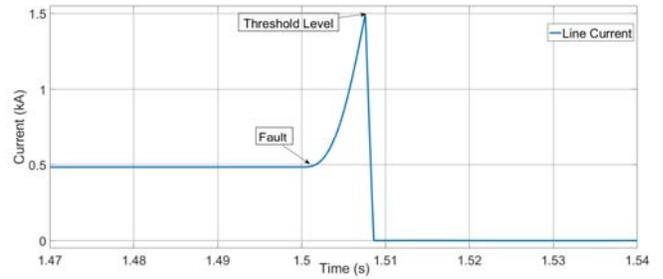

Fig. 12: Fault current in DC link.

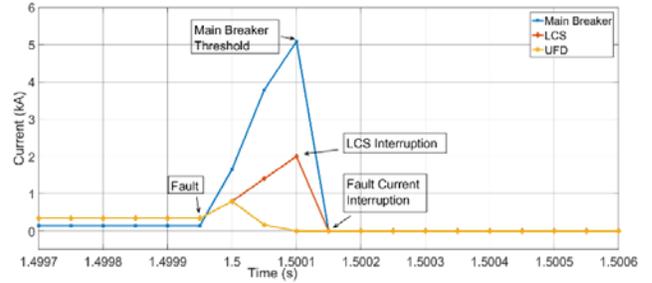

Fig. 13: Hybrid DC breaker current characteristics.

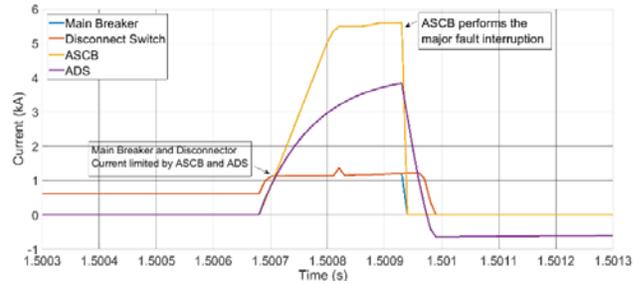

Fig. 14: Assembly HVDC breaker current characteristics.

Table II:
Comparative Analysis of DC breakers

|  | Hybrid DC Breaker | Assembly HVDC Breaker |
|---|---|---|
| Current Breaking Capability | 3 kA -6 kA | 3 kA - 9 kA |
| Major Investment | Main Breaker (MB) | Active Short Circuit Breaker (ASCB) |
| Speed of Operation | Similar | Similar |
| Components required | Two per line | Two breakers per line, and one ASCB per converter |
| Investment Cost | Higher, due to more number of Main Breaker Units | Lower, less number of ASCB units required for interruption |
| Losses | Negligible | Negligible |
| Converter Performance | Unaffected | Unaffected |

As the ASCB is turned on at the converter buses, it creates a temporary short circuit condition at the DC bus, but this action lasts for a few milliseconds, and thus, does not hamper the stability of the network. Even if the DC bus is weak, it does not hamper the ride through capability of the overall network.

The ADS has a discharge resistor connected in series, which is selected to be a medium range resistor since a large value of the resistance will give rise to a large voltage stress that will appear across the IGBT modules. It is mainly a protection unit designed to reduce the voltage stress across the main breaker when opened.

## VIII. Conclusion

In this paper, methods to detect and clear the DC side fault in VSC HVDC systems has been discussed using two different designs of DC breakers. The main breaker unit of the hybrid DC breaker does the major fault current interruption. To maintain proper operation, two hybrid DC breaker units are needed to be placed in every line. Also since the fault current flows through the LCS and UFD for a period of time before being turned off, they have to be designed at a capacity higher than normal rating. The major investment in designing a hybrid breaker is where the major fault current interruption takes place.

On the other hand, assembly HVDC breakers require only one major current interruption unit placed at the common DC bus of the converter stations. So the total number of major current interruption units required would be similar to the number of converter stations present. This is a more cost effective strategy owing to the fact that assembly HVDC breaker offers the same operational characteristics with higher current interruption capabilities.

A three terminal bipolar VSC HVDC link was designed along with its associated protection system using PSCAD/EMTDC. Simulation results show an advantage of using the assembly HVDC breaker over the conventional hybrid DC breaker in meshed HVDC networks.

## Acknowledgment

The authors would like to acknowledge the Energy Production and Infrastructure Centre (EPIC) of that University of North Carolina at Charlotte for their support.

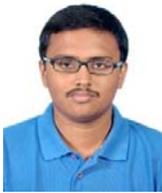

**Bhaskar Mitra** (StM' 15) received his B.Tech degree in Electrical Engineering in 2013 from West Bengal University of Technology. He is currently a Ph.D. student in the Electrical & Computer Engineering department of the University of North Carolina at Charlotte working under Dr. Badrul Chowdhury. His research interests include Protection and Control of Multi-terminal HVDC and also developing new protection coordination and fault localization schemes for MT-HVDC systems.

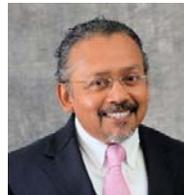

**Badrul Chowdhury** (StM'83, M '87, SM '93) obtained his B.S degree from Bangladesh University of Engineering & Technology in 1981, his M.S. and Ph.D. degrees from Virginia Tech, Blacksburg, VA in 1983 and 1987 respectively, all in Electrical Engineering. He is currently a Professor in the Electrical & Computer Engineering department of the University of North Carolina at Charlotte. Prior to joining UNC-Charlotte, he was a Professor in the Electrical & Computer Engineering department of the Missouri University of Science and Technology, formerly known as the University of Missouri-Rolla. Dr. Chowdhury's research interests are in power system modeling, analysis and control, and renewable and distributed energy resource modeling and integration in smart grids. He is a Senior Member of the IEEE.